\documentclass{emulateapj}

\font\tenbg=cmmib10 at 10pt
\def \rvecmu{{\hbox{\tenbg\char'026}}}
\def \rvecphi{{\hbox{\tenbg\char'036}}}

\font\tenbg=cmmib10 at 10pt

\def \rvecphi{{\hbox{\tenbg\char'036}}}

\def \rvecomega {{\hbox{\tenbg\char'041}}}

\begin{document}

\title{Winds, B-Fields, and Magnetotails
of  Pulsars}

\author{M. M. Romanova, G. A. Chulsky, and
R. V. E. Lovelace} \affil{Cornell University,
Ithaca, NY 14853}

\begin{abstract}

We investigate the emission of rotating magnetized neutron stars
due to the acceleration and radiation of particles in the
relativistic wind and in the magnetotail of the star.
    We consider
that the charged particles are accelerated by driven collisionless
reconnection. Outside of the light cylinder, the star's rotation
acts to wind up the magnetic field to form a predominantly
azimuthal, slowly decreasing with distance, magnetic field of
opposite polarity on either side of the equatorial plane normal to
the star's rotation axis. The magnetic field annihilates across
the equatorial plane with the magnetic energy going to accelerate
the charged particles to relativistic energies.
  For a typical
supersonically moving pulsar, the star's wind extends outward to
the standoff distance with the interstellar medium.
   At larger
distances, the power output of pulsar's wind $\dot{E}_w$ of
electromagnetic field and relativistic particles is {\it
redirected and collimated into the  magnetotail} of the star.
  In the magnetotail it is proposed
that equipartition is reached
between the magnetic energy
and the relativistic particle energy.
     For such conditions, synchrotron radiation from the magnetotails
may be a significant fraction of $\dot{E}_w$ for high velocity
pulsars. An equation is derived for the radius of the magnetotail
$r_m(z^\prime)$ as a function of distance $z^\prime$ from the
star. For large distances $z^\prime$, of the order of the distance
travelled by the star, we argue that the magnetotail has a
`trumpet' shape owing to the slowing down of the magnetotail flow.
We compare results with the Guitar and Mouse nebula, and conclude
that the  tail of the Mouse may be connected with the long
magnetotail behind the pulsar. We argue that the shock waves and
elongated structures may
%%%%%%%%%%%%%%% mmm %%%%%%%%%%%
also
%%%%%%%%%%%%%%%%%
be observed in a misdirected or shutoff pulsars and may be
used as a tool for finding these objects.

\end{abstract}


\section{Introduction}

     The relativistic winds of supersonically moving pulsars are
observed to form shock waves in the interstellar medium (ISM) and
in some cases prominent wakes or ``tails".
       Several pulsars are known to
power  bow shock waves and tails, the ``Guitar'' nebula (Cordes et
al. 1993; Chatterjee \& Cordes 2002), millisecond pulsars
(Kulkarni \& Hester 1988; Bell et al. 1995),   the ``Duck'' nebula
(e.g., Thorsett, Brisken, \& Goss 2002), the ``Mouse'' nebula,
observed in the X-ray and radio bands  and powered by the
spin-down of a $98$ ms pulsar (Camilo et al. 2002; Gaensler et al.
2004), the pulsar B0740-28 (Jones, Stappers, \& Gaensler 2002),
and the neutron star candidate RX J1956.5-3754 (van Kerkwijk \&
Kulkarni 2001a, b).
Axisymmetric simulations of the interaction of  magnetized stars
with the interstellar medium (ISM) have shown that such
interaction leads to formation of long magnetotails  behind the
stars (Romanova et al. 2001;  Toropina et al. 2001).
   Such magnetotails may allow  relativistic
particles from the pulsar to propagate to
large distance behind the pulsar.

        In isolated pulsars a
relativistic magnetohydrodynamic (MHD) wind forms outside of the
light cylinder and propagates outward (Goldreich \& Julian 1969,
hereafter GJ69; Arons \& Tavani 1994; Arons 2004). This wind forms
a magnetically dominated, azimuthally wrapped disk-like equatorial
structure  which expands with velocities
of the order of the speed of light
and is driven mostly by the magnetic pressure.
Observations reveal disk-like structures around a number of
pulsars, including one in the Crab nebula (Weisskopf, et al.
2000), the Vela pulsar (Pavlov et al. 2001), and a few other cases
(Helfand, Gotthelf, \& Halpern 2001; Gotthelf 2001).
     The orientation
and thickness of these disk-like structures (or, PWN tori)
   were  estimated by  Ng and Romani (2004) who
suggested that particles are accelerated more efficiently in these
disk-like structures.
It was also noticed that the rotation axis of the disk
approximately coincides with the direction of pulsar's velocity,
so that the geometry of the flow is approximately axi-symmetric
(Spruit \& Phinney 1998; Lai, Chernoff, \& Cordes 2001). In a few
cases  jets were observed  which are approximately aligned with
the direction of motion (e.g., Weisskopf, et al. 2000).

    Formation of such an equatorial disk
structure was recently modelled in axisymmetric relativistic MHD
simulations by Del Zanna, Amato, and Bucciantini (2004), and
Komissarov  and Lyubarsky (2004) for the case of an intial
split monopole magnetic field.
     They have shown that
a magnetically-dominated disk is formed around a rotating
    pulsar and that this disk has a small thickness.
Formation of rotating, azimuthally wrapped disk-like structures
were also observed in non-relativistic
axisymmetric simulations of an accreting, rotating star with
an aligned dipole magnetic field in the
propeller regime (Romanova et al. 2003).
    This disk-like magnetically dominated bulk flow carries most of
energy out of the pulsar, however, it may be invisible unless the
particles are accelerated by some mechanism.

        Different models have been proposed to
explain acceleration of particles in this wind.
        In one class of models {\it ideal} relativistic
magnetohydrodynamics is assumed to hold outside of the light
cylinder with the acceleration of plasma due to the spatial
variation of the electric and magnetic fields (e.g., Vlahakis
2004).
        In these models it is suggested that
the Poynting-flux to matter-energy-flux ratio ($\sigma$) evolves
from a value much larger than unity near the light cylinder to a
value less than unity at large distances where the pulsar wind
encounters the interstellar medium.
         In another class of models {\it non-ideal}
plasma effects are considered to be responsible for the particle
acceleration.
        The particle acceleration may be stochastic
as discussed  in the shock wave acceleration models (Arons \&
Tavani 1994; Arons 2004; Spitkovsky \& Arons 2004).
        Alternatively in collisionless
reconnection and annihilation the magnetic field may cause random
(Coroniti 1990) or bulk acceleration of the particles (Lyubarsky
\& Kirk 2001; Kirk \& Lyubarsky 2001). These models argue that the
MHD wind consists of an azimuthally wrapped dipole magnetic field
of the star with regions where the magnetic field reverses
direction.
         These regions are
likely sites for reconnection and particle acceleration.

In this work we investigate a model where particles are
accelerated in the pulsar's {\it equatorial}
neutral layer across which
the predominantly azimuthal magnetic field reverses direction.
       There may also be oppositely directed
Poynting flux flows aligned with the star's rotation axis
analogous to those generated
from the time-dependent expansion of
magnetic field loops of an accretion disk into a low density
coronal plasma
discussed  by Lovelace and Romanova (2003)
and simulated by Lovelace,
Gandhi, and Romanova (2004) with a relativistic
electromagnetic  particle-in-cell
method.
     However, note that numerical solutions of the ``pulsar
equation'' for an aligned stationary magnetosphere
   by Contopoulos,
Kazanas, and Fendt (1999) and Gruzinov (2005)
  do not show Poynting flows along the rotation axis.

        Our model is  qualitative in
that we do not have global self-consistent calculations of the
electromagnetic fields and sources.
       Compared to Coroniti (1990) we suppose
that the misalignment angle between the rotational and magnetic
axes
%%%%%%%%%%%%%%%RRR%%%%%%%%%%%%%%%%%%%%%%
{\it is not large}. Then, instead of multiple neutral layers in a
system of magnetic stripes, a global extended  neutral layer forms
in the equatorial plane.
    The equatorial layer
is not expected to be  thin everywhere.
    However, the systematic acceleration will
be in the radial direction.
     Thus the bulk motion of plasma and magnetic field
will be in the radial direction mainly in the equatorial plane.
   Thus, we expect to have a
high-$\sigma$ plasma near the light cylinder which
   gradually changes to a
much lower $\sigma$  far from the pulsar.
    Accelerated particles will gradually
diffuse vertically away from the
thin neutral layer, and their interaction
with the magnetic field
will produce the synchrotron
radiation.
    In this paper we develop this model
in semi-quantitative detail.

       The acceleration of particles in
collisionless reconnection has been discussed by a number of
authors (Alfv\'en 1968; Dessler 1969, 1971; Speiser 1970; Cowley
1971, 1973; Bulanov \& Sasorov 1976; and Vasyliunas 1980;
Burkhart, Drake, \& Chen 1991).
       The present work uses
the model by Alfv\'en (1968) and its relativistic counterpart
(Romanova \& Lovelace 1992; Zenitani \& Hoshino 2001; Larrabee,
Lovelace, \& Romanova 2003).  Both the simulations
(Zenitani \& Hoshino 2001) and the analytic theory
(Larrabee et al. 2003) indicate a power law distribution
of accelerated particles.

       Most  pulsars propagate
through the ISM  highly supersonically
so that the pulsar wind forms a
strong shock wave in the ISM.
        A model of this
interaction was first discussed by Schvartsman (1970).
        Subsequently,  the
shape of the shock wave
was calculated for the hydrodynamic flow
by Baranov, Krasnobaev \& Kulikovsky
(1971), Baranov \& Malama (1993) and  by Wilkin (1996).
       However, for the pulsar case it
is necessary to include the relativistic
magnetized wind of the star.
     Recent models of the Solar wind
interaction with the ISM
include the influence of the
magnetic field for both
the interstellar and interplanetary
components, and these models
show that the shape of the bow shock may
depend on these fields (e.g., Linde et al. 1998;
Zank 1999; Bucciantini et
al. 2004; Opher et al. 2004).
   However, only relatively low Mach numbers are
considered and the Alfv\'en speeds are typically
small compared with the flow speeds.
      In the case of pulsars
the Mach numbers are much larger than unity and
the Alfv\'en speeds may be comparable with
the flow speeds.
    Strongly magnetized magnetotails were observed
in axisymmetric simulations of rotating and non-rotating
strongly magnetized stars (Toropina et al. 2001).

  In this paper we discuss the nature of
the magnetotails which  form
behind the pulsars.
We argue that most of the power
of the star's wind of relativistic
particles and magnetic field
is {\it redirected} by the internal
shock wave and subsequently
propagates down the magnetotail.
       In several cases observations show cometary
structures aligned with the
direction of propagation of the pulsar
(e.g., Stappers et al. 2003;
      Gaensler et al. 2003)
which may be the result of the
pulsar wind propagating down the
magnetotail.

Both shocks and magnetotails
may appear in the case of misdirected
pulsars or longer-period, shutoff pulsars.
We discuss the possible
nature of the isolated neutron
star candidate RX J1956.5-3754
which has a  shock wave.

In \S 2 we discuss basic considerations.
   In \S 3 we discuss
particle acceleration due to
reconnection in the equatorial
neutral layer of the star,
and in \S 4 the associated synchrotron
radiation.
   In \S 5 we discuss the
bow shock wave and the related
magnetotail and its synchrotron emission.
   In \S 6 we discuss examples of
pulsars with tails.
      In  \S 7 we consider the
possibility of observing wind
and magnetotail emitting stars in
the Solar neighborhood.
    In \S 8 we give the conclusions of this
work.

\section{Basic Considerations}

%%%%%%%%%%%%%%%%%%%%%%%%%%%%%%%%%%%%%%%%%%%%%%%%%%%%
\begin{figure*}[t]
\epsscale{1.0} \plotone{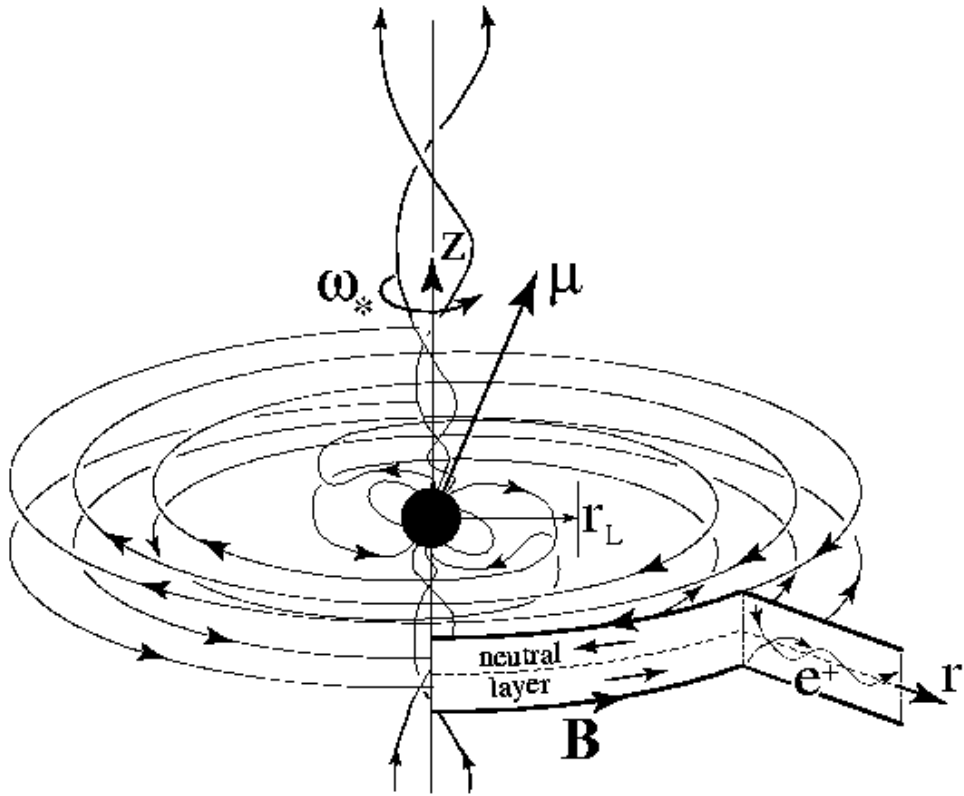} \caption{Sketch of the magnetic
field configuration of misaligned rotating neutron star outside of
the light cylinder, $r > r_L =c/\omega_*$. For the case shown
${\bf \Omega}\cdot \rvecmu >0$ the magnetic field is wrapped in
clock-wise spiral in the upper half-space and a counter-clockwise
spiral in the lower half-space. A neutral layer is formed in the
equatorial plane where there is forced reconnection or
annihilation of the magnetic field. The magnetic field energy goes
into accelerating particles to relativistic energies.
      The helical field lines around the $\pm z$ axes
correspond to  possible Poynting outflows. }
\end{figure*}
%%%%%%%%%%%%%%%%%%%%%%%%%%%%%%%%%%%%%%%%%%%%%%%%%%%

A pulsar's light cylinder radius is
\begin{equation}
r_L={c\over \omega_*}={P c \over 2\pi}\approx4.8\times10^9~P~{\rm
cm}~,
\end{equation}
where $P$ is the period in seconds. The pulsar spins down owing to
the relativistic wind which flows outward beyond the light
cylinder distance (GJ69). This wind consists of an electromagnetic
field and relativistic particles and has a power output
\begin{equation}
\dot{E}_{w} \approx B_L^2 r_L^2 c ={\mu^2 \omega_*^4 \over c^3}
\approx 5.8\times 10^{31} {\mu_{30}^2\over P^4}~ {\rm{ erg\over
s}}~,
\end{equation}
(GJ69), where $B_L \equiv \mu/r_L^3$ is the magnetic field
strength at the light cylinder radius, and $\mu_{30} =
\mu/(10^{30}{\rm G cm}^3)$ is the star's magnetic moment.
        The star's surface magnetic field is
$B = B_{12}10^{12}{\rm G}$, with $B_{12} = \mu_{30}/R_6$, where
$R_6=R_*/10^6{\rm cm}$ and $R_*$ the star's radius.

Assuming the pulsar wind approximately isotropic, the pressure of
the wind varies as $p_w = \dot{E}_w/(4\pi R^2 c)$ for distances $R
>   r_L$ from the star. For a pulsar moving supersonically through
the ISM, the pressure of the pulsar wind $p_w(z)$ balances the ram
pressure of the ISM $\rho_{ism}v^2$ at the stagnation point at a
distance $z_{sh}$ in front of the pulsar. This is the location of
the bow shock of the pulsar. That is, $p_w(z_{sh})=\rho_{ism}v^2 $
so that
\begin{equation}
z_{sh}={B_L r_L \over (4\pi \rho_{ism}v^2)^{1/2}} \approx 10^{14}
{B_{12} \over n_{ism}^{1/2} ~v_8~P^2} ~{\rm cm}~,
\end{equation}
(Schvartsman 1970; Baranov et al. 1971;
Wilkin 1996), where $\rho_{ism}/m_p\approx n_{ism}$ (in units of
cm$^{-3}$) is the number density of the ISM with $m_p$ the proton
mass.
        The lateral width of the bow shock is $ \approx
z_{sh}$. Thus the power which goes into {\it thermal heating} of
the shocked interstellar gas is
\begin{equation}
\dot{E}_{sh} \approx {3\over 16} \rho_{ism} v^3 \pi z_{sh}^2
\approx 0.9\times 10^{28}~ {v_8 B_{12}^2 \over P^4}~{\rm erg \over
s},
\end{equation}
where the $3/16$ factor assumes a specific heat ratio of $5/3$.
The power $\dot{E}_{sh}$ is extracted from the kinetic energy of
the motion of the star through the ISM. The slowing of the star's
motion is however negligible. Notice that the ratio
\begin{equation}
{\dot{E}_{sh} \over \dot{E}_w} \approx {3\over 64} {v \over c}
\approx 1.56 \times 10^{-4}~ v_8~,
\end{equation}
is much smaller than unity.

%%%%%%%%%%%%%%%%%%%%%%%%%%%%%%%%%%%%%%%%%%%%%%%%%
\begin{figure*}[t]
\epsscale{1.0} \epsscale{1}\plotone{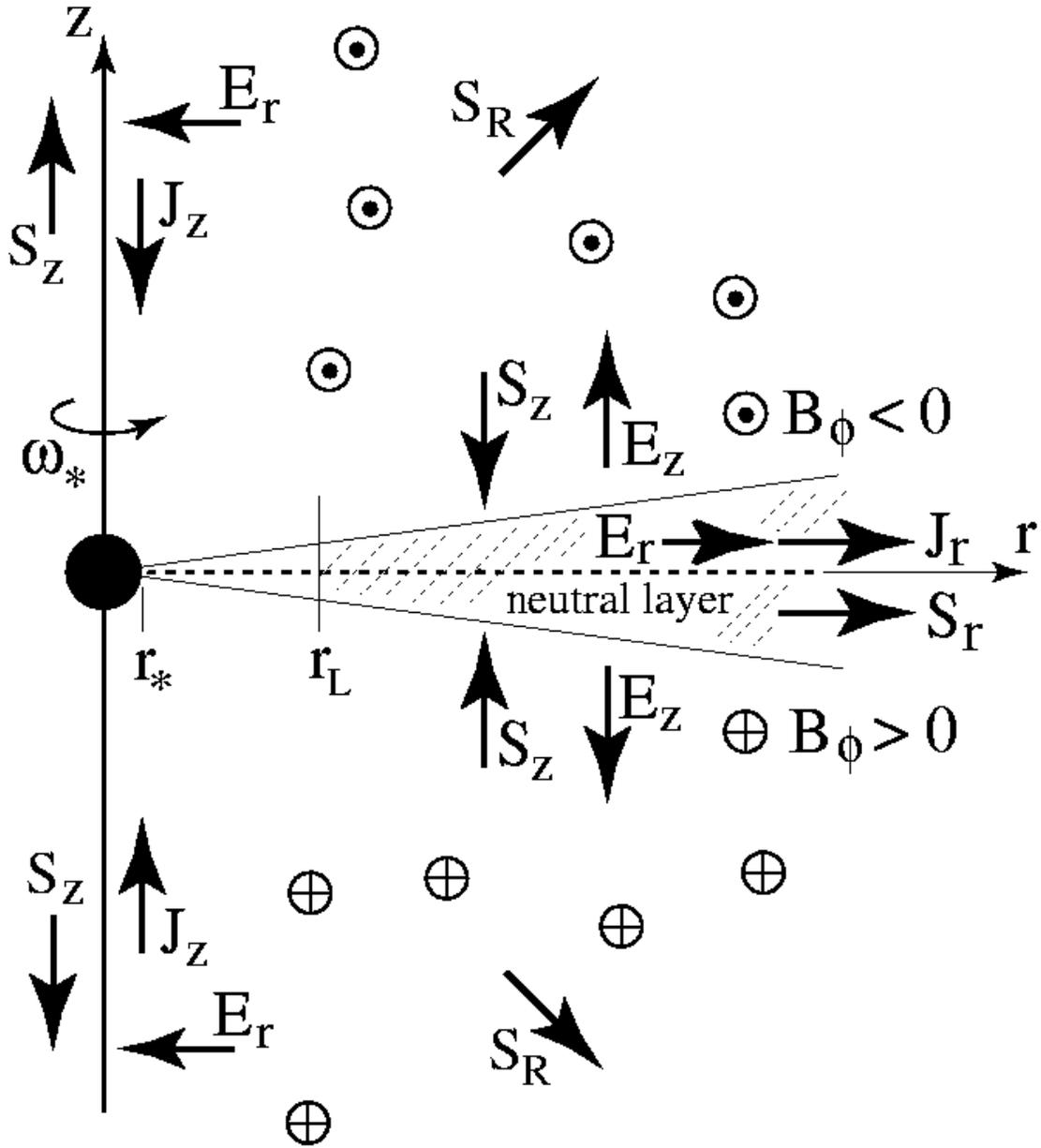} \caption{The figure
shows the current flows (${\bf J}$), electric fields (${\bf E}$),
and Poynting vectors (${\bf S} =c{\bf E \times B}/4\pi$) of the
envisioned configuration outside of the light cylinder, $r> r_L =
c/\omega_*$. }
\end{figure*}
%%%%%%%%%%%%%%%%%%%%%%%%%%%%%%%%%%%%%%%%%%%%%%%%%

\section{Magnetic Reconnection}

Outside of the light cylinder the pulsar's magnetic field is
predominantly azimuthal. If the magnetic moment of the star is
perpendicular to its rotation axis, then the azimuthal magnetic
field forms ``stripes" of opposite polarity which move outward
relativistically (e.g., Michel 1971, Coroniti 1990, Lyubarsky \&
Kirk 2001). Particles may be gradually accelerated by reconnection
of the oppositely directed fields (Coroniti 1990).
        Most of the power flow is
thought to be in the form of a magnetically dominated wind. That
is, the ratio of the magnetic energy density to the kinetic energy
density of the particles, the $\sigma$ parameter, is  much larger
than unity.
        As the distance from the light
cylinder increases, $\sigma$  decreases.

     We consider the case where the angle between the
magnetic and rotation axes is not very large.
       The nature of the envisioned driving mechanism is
sketched in Figures 1 and 2.
       This work focuses on the equatorial region
of the flow, but for completeness the figures
show possible Poynting outflows
along the $\pm z$ axes.
        The plasma  behavior is
{\it non-ideal};  it is {\it not} described by ideal MHD.
        We use both cylindrical $(r,\phi,z)$
and spherical $(R,\theta, \phi)$ non-rotating coordinates to
describe the electromagnetic field and particle motion for
$R>r_L$. The rotation of the star wraps the field around the
rotation axis so that it is predominantly azimuthal as shown in
Figure 1. For the case assumed here, ${\rvecomega}_*\cdot \rvecmu
>0$, the field lines are wound up clockwise for $z>0$ and
counter-clockwise for $z<0$. Consequently, there is a {\it neutral
layer} in equatorial plane where the azimuthal magnetic field
changes from $B_\phi >0$ for $z<0$ to $B_\phi <0$ for $z>0$.
Within the neutral layer, the magnetic field is approximated as
\begin{equation}
B_r=B_L\left({r_L\over r}\right)^2 \left({z\over \Delta
z}\right)~,\quad B_\phi = -B_L \left({r_L \over
r}\right)\left({z\over \Delta z}\right)~,
\end{equation}
where $\Delta z(r)$ is the half-thickness of the neutral layer
which is assumed less than $r$,
     and where $B_L \equiv \mu/r_L^3
\approx 9.2~\! \mu_{30}/P^3$ G. Thus the magnetic field has form
of Archimedes' spiral with field lines given by $r={\rm
const}-r_L\phi$.
Equation (6) neglects the effect of the annihilation of the field.

%%%%%%%%%%%%%%%%%%%%%%%%%%%%%%%%%%%%%%%%%%%%%%%%%
\begin{figure*}[t]
\epsscale{1.} \plotone{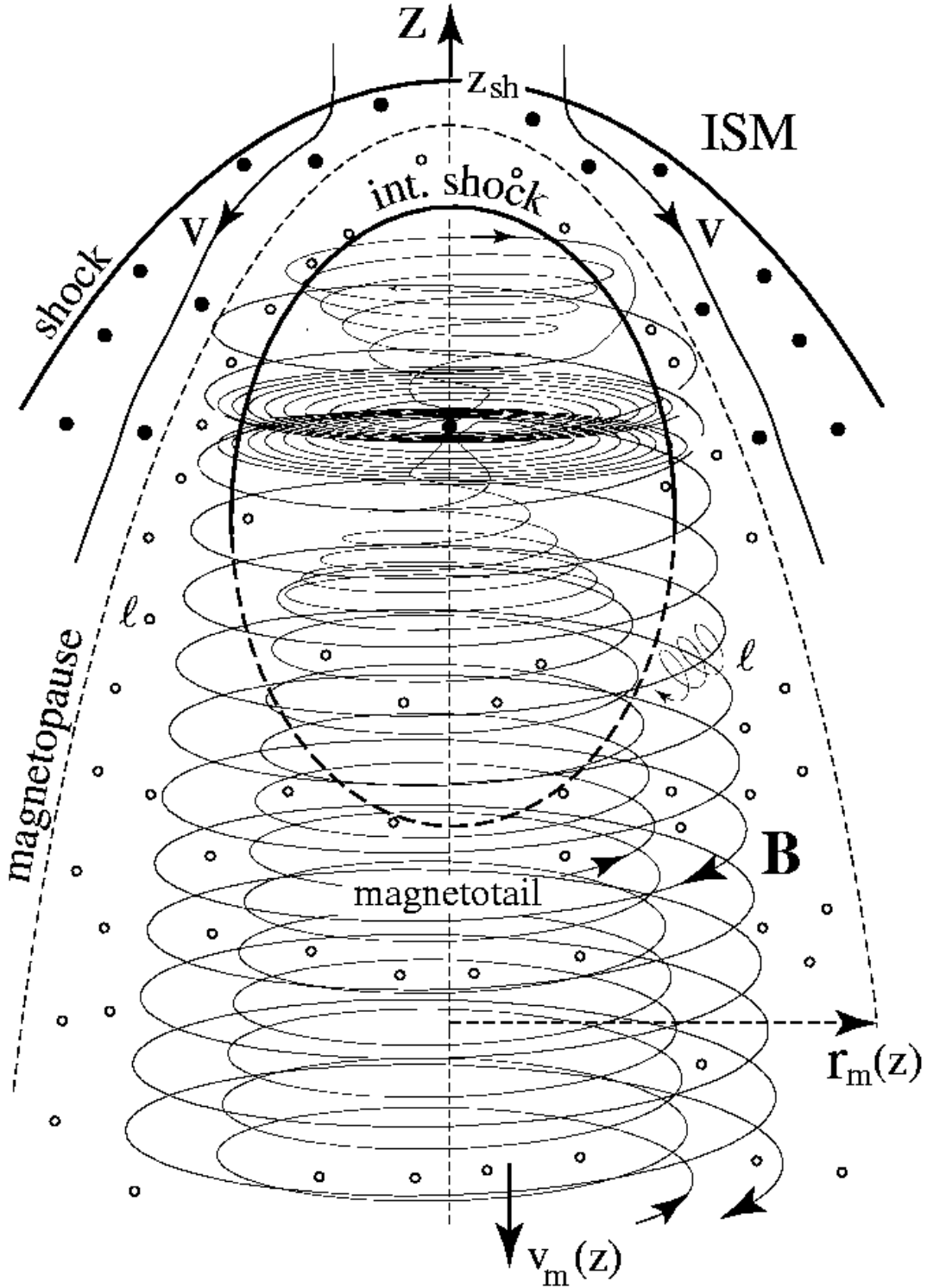} \caption{Sketch of the external
shock, the magnetopause, and the magnetotail of a pulsar moving
supersonically through the interstellar medium. The velocity of
the star is assumed parallel to the star's rotation axis
$\rvecomega_*$. The solid circles indicate the shocked
interstellar gas with sample streamlines labelled by ${\bf v}$.
The open circles indicate relativistic leptons ($\ell$) which flow
into the magnetotail. The radius of the magnetotail is $r_m(z)$
and the bulk velocity of the magnetotail plasma is ${\bf v}_m(z)$.
        The internal shock of the pulsar wind
is indicated by ``int. shock.'' The standoff distance of the
shock, $z_{sh}$, is given by equation (3). }
\end{figure*}
%%%%%%%%%%%%%%%%%%%%%%%%%%%%%%%%%%%%%%%%%%%%%%%%%

In the neutral layer Amp\`ere's law gives $\partial
B_\phi/\partial z =-4\pi J_r/c$. Thus the change in the azimuthal
field across the layer is
\begin{equation}
\Delta B_\phi = \bigg[B_\phi\bigg]_{-\Delta z}^{\Delta z} = -~{2
I_r \over c~ r}~.
\end{equation}
Thus, the radial current carried by the neutral layer is
\begin{equation}
I_r = B_L r_L c \approx 4.4 \times 10^{11}~ {\mu_{30} \over
P^2}~{\rm A}~,
\end{equation}
where $\mu_{30} \equiv \mu/(10^{30}{\rm ~G~ cm}^3)$ is the
magnetic moment of the star. This radial equatorial current flow
acts to {\it magnetically pinch} the particles within the neutral
layer. An analogous current layer exists in the solar wind (e.g.,
Bertin \& Coppi 1985).

   The equatorial outflow of current is balanced by an axial inflow
of current as indicated in Figure 2.
   This current corresponds to a
particle outflow rate of
\begin{equation}
{d N \over dt} \approx {I_r \over e} \approx 2.7 \times
10^{30}~{\mu_{30} \over P^2} ~ {\rm s}^{-1}~.
\end{equation}
There may be oppositely directed
Poynting flux outflows along the rotation axis with $S_z =
(c/4\pi) E_r B_\phi~ {\buildrel > \over <}~0$ for $z~{\buildrel > \over
<}~0$ flow outward along the star's rotation axis.  These
outflows would be analogous to the Poynting outflows from
accretion disks
  (Lovelace \& Romanova 2003; Lovelace et al. 2005).

For the considered case ${\bf \Omega}\cdot \rvecmu >0$, the
electric charge of the neutral layer is dominantly positive. Thus
the charge of the layer gives rise to an outward radial electric
field $E_r >0$ and an axial field $E_z >0$ for $z>0$ and $E_z <0$
for $z<0$ as shown in Figure 2. The electric field gives rise to
an axial ${\bf E \times B} =E_r B_\phi{\hat{\bf z}}$ ($<0$ for
$z>0$ and $>0$ for $z<0$) drift of the azimuthal field and
associated plasma {\it into} the neutral layer. This constitutes
the driving force for the reconnection.

Additionally, there is a radially outward ${\bf E \times B}=-E_z
B_\phi \hat{\bf r}$ drift of the particles in the neutral layer.
The ${\bf E \times B}$ drifts are of course in the same direction
as the Poynting vector ${\bf S} = (c/4\pi){\bf E \times B}$ which
is shown in Figure 2.

%%%%%%%%%%%%%%%%%%%%%%%%%%%%%%%%%%%%%%%%%%%%%%%%%%%%%%%
      Particles are accelerated in
the electric field $E_r$ which decreases with the distance as
$1/r$ or faster.
      This dependence is determined by
the fact that the energy comes from the  magnetic field,
     $\sim B_\phi$, which varies as
$1/r$ (see equation 6).
      There is an additional factor in equation (6),
$z/{\Delta z}$, which takes into account the fact that at larger
distances the relative thickness of the neutral layer will likely
increase due to instabilities and turbulence.
     The instabilities may be
driven by the misalignment of the rotation ${\bf \Omega}$ and
magnetic $\rvecmu$ axes.
      This is expected to lead to faster than $1/r$
decrease of the field with the distance.
       In addition, part of the
energy of the field will go to acceleration of particles, which
will cause $E_r$ to fall off more rapidly than $1/r$.
      Thus, we
introduce a parameterized dependence of the electric field within
the neutral layer in the form
%%%%%%%%%%%%%%%%%%%%%%%%%%%%%%%%%%%%%%%%%%%%%%%%%%%%
\begin{equation}
E_r = \alpha_E B_L \left( {r_L \over r}\right)^q~,
\end{equation}
where $\alpha_E$ is a dimensionless quantity assumed to be a
constant less than unity,
%%%%%%%%%%%%%%%%%%%%%%%%%%%%%%%%%%%%%%%%%%%%%%%%%%%%%%
and $q$ is another constant which characterizes how fast this
electric field decreases with the distance.
     The overall electrical neutrality of the equatorial and
axial regions  implies that $q>1$. At the boundary of the neutral
layer $E_z(r,\Delta z) \approx \alpha_E^\prime B_L (r_L/r)$, where
$\alpha_E^\prime$ is further dimensionless quantity assumed to be
appreciably less than unity.

Neglecting for the moment radiative energy losses, positively
charged particles drifting into the neutral layer are accelerated
in the direction $\hat{\bf r} + (r_L/r)\hat{\rvecphi}$ which is
approximately radial for $r \gg r_L$. The radial acceleration
gives
\begin{equation}
{d \over dt}( m c^2 \gamma) \approx c {d \over dr}(m c^2 \gamma )
\approx q \alpha_E c B_L \left( r_L \over r\right)^q~,
\end{equation}
where $m$ is the particle rest mass and $e$ is its charge
(Alfv\'en 1968). The approximation involves assuming that the
particles move outward with speed $\approx c$.
%%%%%%%%%%%%%%%%%%%%%%%%%%%%%%%%%%%%%%%%%%%%
Integrating eq. (11) from $r=r_L$ to $r >> r_L$, we find
%%%%%%%%%%%%%%%%%%%%%%%%%%%%%%%%%%%%%%%%%%%%
\begin{equation}
\gamma = 1 + {e B_L r_L\over m c^2}{ \alpha_E \over (q-1)}~,
\end{equation}
for $r\gg r_L$ and $q$ not close to unity.
\begin{equation}
\gamma \approx 2.6\times 10^7 \left({\alpha_E \over q-1}\right)
\left({\mu_{30} \over P^2}\right)~,
\end{equation}
for the case of leptons. For protons and heavier ions the
numerical factor is $\lesssim 1.4\times 10^4$. A smaller number of
negatively charged particles is accelerated back towards the
pulsar.

The outward kinetic energy flux of the particles accelerated in
the neutral layer is
$$
\dot{E}_{kin} = mc^2(\gamma-1){dN \over dt} =(B_L^2 r_L^2 c)
{\alpha_E \over q-1}$$
\begin{equation}
\approx 5.8\times 10^{30} \left({\alpha_E \over 0.1}\right)
{\mu_{30}^2 \over P^4}~{\rm erg \over s}
\end{equation}
for $r\gg r_L$, neglecting the synchrotron losses. This kinetic
energy flux is a fraction $\alpha_E/(q-1)$ of the Goldreich-Julian
power ($B_L^2 r_L^2 c$).

%%%%%%%%%%%%%%%%%%%%%%%%%%%%%%%%%%%%%%%%%%%%%%%%
     We expect  different behaviors for the case
where $q$ is appreciably larger than unity ($q-1\sim 1$),  and the
case when $q$ is  close to unity ($q-1 \ll 1$).
      In the first case particles are
accelerate mainly in the vicinity of the light cylinder.
       In the
second case, they are accelerated along the neutral layer out to
much larger distances.
      The total energy, however, can
not be larger than the rotational energy of the pulsar, that is
$\alpha_E/{q-1} \leq 1$.

      Particles, accelerated in
the neutral layer will be gradually scattered by irregularities
(Alfv\'en waves) in the magnetic field and will drift away from
the neutral layer.
       At the same time
there may be magnetic  confinement of particle to the neutral
layer out to large distances.
      The force balance (or magnetic pinch) condition for the
neutral layer follows from the momentum equation $\partial
T_{zz}/\partial z=0$ or $T_{zz}=$ const (Bertin \& Coppi 1985),
where $T_{jk}$ is the stress tensor for the particles and fields.
This condition implies that $\alpha_E = (\Delta z/r)
[1-(\alpha_E^\prime)^2]$.
         The compressive pinch force on
the neutral layer is responsible for the reconnection at $z=0$
being driven.

         With increasing distance from the light
cylinder the ratio of the
Poynting to particle energy fluxes
$\sigma$ decreases.
The variation of $\sigma$ with distance may be anisotropic, with
smaller $\sigma$ in the disk and larger $\sigma$ above and below
the disk.
      With increasing $r$, a larger fraction of the
magnetic energy is converted to particle energy in the neutral
layer.
      Particles accelerated in the
neutral layer may subsequently scatter so as to give a more
spherical wind.
       The  scattering and diffusion processes
of particles accelerated in the neutral layer remains to be
investigated in detail.
       For simplicity of the following
analysis we suppose that $\sigma \lesssim 1$ before the bow shock
is reached so that the flow is super-Alfv\'enic and supersonic at
the bow shock.

\section{Synchrotron Radiation}

The acceleration of leptons in the neutral layer is inhibited due
to the energy lost to synchrotron radiation. Including the
synchrotron losses gives
     \begin{equation}
{d \gamma \over d \tilde{r}}= {\gamma_0 \over \tilde{r}^q} - {k
~\gamma^2 \over \tilde{r}^2}~.
\end{equation}
Here, $\tilde{r}\equiv r/r_L$, $\gamma_0 \equiv e\alpha_E B_L
r_L/(mc^2)$, $k\equiv (2/3)\beta_RB_L^2 r_L r_e^2/(m c^2)$, with
$r_e=e^2/(mc^2)$ the classical electron radius, and $\beta_R<1$
accounts for the fact that the magnetic field close to the neutral
layer is less than $B_L/\tilde{r}$. We assume that $\gamma \sim 1$
at the light cylinder distance $\tilde r =1$. The synchrotron
radiation of ions is negligible.
       Equation (15) represents a simplified
model of the synchrotron losses from the neutral layer in that it
neglects the fact that the leptons tend to be focused to the $z=0$
plane where the magnetic field is appreciably reduced.
       However, this
focusing will be offset if as expected there are significant
irregularities (turbulence)  in the plasma and the magnetic field
as in the Solar wind.

For simplicity we examine the case where $q=2$. It is clear that
the Lorentz factors will not exceed
$$
\gamma_{max} = \left({\gamma_0 \over k}\right)^{1/2}=
\left({3e\alpha_E r_L^3 \over 2 \beta_R r_e^2\mu}\right)^{1/2}
$$
\begin{equation}
\approx 3.1\times 10^7 \left({\alpha_E \over \beta_R}\right)^{1/2}
{P^{3/2} \over \mu_{30}^{1/2} }~.
\end{equation}
Thus the synchrotron losses are significant for pulsar periods $P$
shorter than the period which gives $\gamma_{max}=\gamma_0$. This
critical period is
\begin{equation}
P_{cr}={2\pi \over c} \left({2 \alpha_E \beta_R r_e^3 \mu^3 \over
3e mc^2 }\right)^{1/7}\!\approx 0.49 \left({\alpha_E\beta_R \over
0.01}\right)^{1/7} \mu_{30}^{3/7}~{\rm s}~.
\end{equation}
For $P<P_{cr}$ the synchrotron energy loss rate is
$$
\dot{E}_{syn} = \int_{r_L}^\infty \!\!dr I_r E_r =\alpha_E B_L^2
r_L^2 c
$$
\begin{equation} \approx
5.8\times 10^{30} \left({\alpha_E\over0.1}\right) {\mu_{30}^2
\over P^4}~{\rm erg \over s}~,
\end{equation}
which is a fraction $\alpha_E$ of the Goldreich-Julian power. For
$P<P_{cr}$ the kinetic energy flux of the particles is
$$
\dot{E}_{kin} = (\gamma_{max}-1)mc^2{dN \over dt}
$$
\begin{equation}
\approx 7\times 10^{31} \left({\alpha_E \over
\beta_R}\right)^{1/2} {\mu_{30}^{1/2} \over P^{1/2}}~{\rm erg
\over s}~.
\end{equation}
For $P<P_{cr}$ we have $\dot{E}_{kin}/\dot{E}_{syn} \approx
12.2(P^{7/2}/\mu_{30}^{3/2}) <1$ for $\alpha_E=0.1=\beta_R$.

For $P>P_{cr}$ the synchrotron losses are small compared with
power input to the particles in the neutral layer. The total
synchrotron radiation from the neutral layer ($r\geq r_L$) is
$$
\dot{E}_{syn} ={2\over 3} r_e^2 r_L \beta_R B_L^2 \gamma_0^2 {dN
\over dt}
$$
\begin{equation}
\approx 3.9\times 10^{29}\beta_R \left({\alpha_E \over
0.1}\right)^2 {\mu_{30}^5 \over P^{11}}~{\rm erg \over s}
\end{equation}
The kinetic energy input to the particles in the neutral layer is
$\alpha_E B_L^2 r_L^2 c$.

The synchrotron radiation of the particles accelerated in the
neutral layer is beamed radially outward in the equatorial plane
of the pulsar. That is, the radiation is in a fan beam
perpendicular to the pulsar's rotation axis.
       In reality, the neutral layer
is probably not a thin layer with small  width $\Delta z$
everywhere.
      The accelerated particles will
scatter  from irregularities in
the neutral layer and will interact with
the magnetic field at larger distances, $z >> \Delta z$.
      This process may explain the
radiation from the disk-like pulsar wind nebulae such as those in
the Crab and Vela nebulae.
%%%%%%%%%%%%%%%%%%%%%%%%%%%%%%%%%%%%%%%%%%%%%%

For $P<P_{cr}$ the frequency of the peak of the radiation spectrum
is
$$
\nu_{syn} \approx {1\over 4\pi} { e B_L \gamma^2 \over m c}
\left({r_L \over r}\right)
$$
\begin{equation}
\approx 1.3\times 10^{22} \left({\alpha_E \over \beta_R}\right)
\left({r_L \over r}\right){\rm Hz}~,
\end{equation}
where the lepton Lorentz factor is from equation (16). This
frequency corresponds to a photon energy of $\approx 54$ MeV. For
$P>P_{cr}$, the Lorentz factor of the leptons is $\gamma_0$ so
that
\begin{equation}
\nu_{syn}\approx 8.5\times 10^{19} \left({\alpha_E \over
0.1}\right)^2 {\mu_{30}^3 \over P^7}~{\rm Hz}~.
\end{equation}
This frequency corresponds to $\approx 350$ keV. These numbers are
of course estimates because the value of $\alpha_E$ is not known.

\section{The Magnetotail}

Beyond the light cylinder the pulsar relativistic wind carries the
power $\dot{E}_w = B_L^2 r_L^2 c$ in the electromagnetic field and
relativistic particles (GJ69). The spin down of the pulsar is
given by $I \omega_* \dot{\omega}_*= -\dot{E}_w$ which implies
that
\begin{equation}
P^2=P_0^2 \left(1+{t\over\tau}\right)~, \quad \tau \equiv { IP_0^2
c^3 \over 8\pi^2 \mu^2 }~,
\end{equation}
where $t$ is the age of the pulsar, $P_0$ is the initial period of
the pulsar, and $\tau \approx 1.4 \times 10^{11} (P_0/0.02{\rm
s})^2/\mu_{30}^2~{\rm s}\approx 4,300 (P_0/0.02{\rm
s})^2/\mu_{30}^2~{\rm yr}$ assuming $I=10^{45}{\rm g~cm}^2$. After
a time $t$ the total energy put out by the pulsar is
$E_w=I\omega_{*0}^2[t/(t+\tau)]/2$, where $\omega_{*0}=2\pi/P_0$.

The pulsar wind also carries toroidal magnetic flux outward. The
rate of transport of positive toroidal flux is
\begin{equation}
\dot{\Phi}_+ =\pi B_L r_L c \approx 4.1\times 10^{21}~ {\mu_{30}
\over P^2}~ {{\rm G~ cm}^2 \over {\rm s}}~.
\end{equation}
After a time $t$ the total (positive) toroidal flux put out by the
pulsar is $\Phi_+ =(\pi\mu \omega_{*0}^2 \tau/c)\ln(1+t/\tau)
=(\pi/2)\mu I c^2 \ln(1+t/\tau)$. Of course the net flux is zero.

For early times after the pulsar's birth there is a spherical
Sedov-Taylor ``bubble'' of plasma of radius ${\cal R}(t)\propto
t^a$ with $a<1$. Here we consider {\it late times} in the sense
that the pulsar has had time to move outside of this bubble. That
is, we consider $t > {\cal R}(t)/v$, where $v$ is the pulsar's
velocity. This corresponds to $t> 3 \times 10^{9}{\rm s} ({\cal
R}_{pc}/v_8)$, where ${\cal R}_{pc}$ is the bubble radius in pc,
and $v_8 \equiv v/10^{8}{\rm cm/s}$.

Figure 3 shows a sketch of the late stage of the magnetotail of a
neutron star moving supersonically with velocity $v$ through the
interstellar medium. For simplicity we have assumed that the
star's velocity is parallel to its rotation axis $\rvecomega_*$.
The standoff distance of the shock $z_{sh}$ is given by equation
(3). The pulsar wind extends out to a radial distance $r_{m0}
\approx z_{sh}$ from the $z-$axis at $z=0$. The magnetic field
strength at this distance is
$$
B_{m0} = B_L \left({ r_L\over r_{m0} }\right) =\big(4\pi
\rho_{ism} v^2~\!\big)^{1/2}
$$
\begin{equation}
\approx 4.6\times 10^{-4}~n_{ism}^{1/2}~ v_8~{\rm G}~,
\end{equation}
which is independent of the period and magnetic moment of the
pulsar. The magnetic field pressure of the wind at $r_{m0}$ is
\begin{equation}
p_{m0}={{\bf B}^2 \over 4\pi}\bigg|_{r_{m0}} = \Gamma~ p_{ism}
{\cal M}^2~,
\end{equation}
where $p_{ism}$ is the ambient pressure of the interstellar
medium, $\Gamma$ is the ratio of specific heats, and ${\cal
M}=v/c_s$ is the Mach number of the pulsar.
       Typically,
${\cal M} \gg 1$, so that electromagnetic pressure at $r_w$ is
much larger than $p_{ism}$. The pressure of the shocked
interstellar gas is larger than $p_{ism}$ by a factor ${\cal M}^2$
neglecting radiative cooling. Because the time scale $t_m \equiv r_{m0}/v$ 
is much shorter than the spin-down time scale $\tau$,
the time-dependence of the pulsar's period can be neglected.

        The momentum of the relativistic pulsar
wind is {\it deflected} by the {\it internal shock} as indicated
in Figure 3.
       The flow of power and magnetic flux
is however  essentially unchanged.
        We assume that the internal shock has
the effect of establishing an {\it approximate equipartition}
between the field energy density and the energy density of
relativistic particles.
       It is unlikely that the pulsar's
magnetized wind is compressed into a thin layer in contact with
the interstellar medium as discussed in the hydrodynamic model
(Baranov, Krasnobaev, \& Kulikovskii 1971; Wilkin 1996). Plasma in
the magnetotail moves away from the pulsar with an initially {\it
relativistic} velocity, ${\bf v}_m \approx v_{mz}\hat{\bf z}$,
where $v_{mz}(z=0) = {\cal O}(c)$. As indicated in Figure 3, the
magnetic field in the magnetotail is predominantly toroidal, ${\bf
B} \approx \hat{\rvecphi~\!}B_m(z)$.

        Inside the magnetotail there is
a cylindrical neutral layer across which the magnetic field
reverses direction (see also simulations by Romanova et al. 2001;
Toropina et al. 2001).
        Reconnection or annihilation
of the magnetic field in this layer continually accelerates the
charged particles of the flow.
        In the following
equations, we assume that the charged particles consist of
electrons and positrons.
       The flux of particles from
the pulsar is $dN_\ell/dt \approx 2B_L r_L c/e$. In steady state
this particle flux flows down the magnetotail so that $dN_\ell/dt
= \pi r_m^2 n_\ell ~v_{mz}$, where $n_\ell$ is the lepton density.
The equipartition Lorentz factor of the leptons is
$$
<{\gamma}_\ell> = {1\over 8}{eB_Lr_L\over mc^2} \left({B_m r_m
\over B_{m0} r_{m0}}\right)^2 {v_{mz} \over c}~,
$$
\begin{equation}
\approx 10^6 ~{\mu_{30} \over P^2} \left({B_m r_m \over B_{m0}
r_{m0}}\right)^2 \left({v_{mz} \over 10^{10}~{\rm cm/s}}\right)~,
\end{equation}
where we have used the fact that $B_{m0}r_{m0}= B_L r_L$.

The total, kinetic plus field, energy flux along the magnetotail
is
\begin{equation}
{\cal F}_E(z) ={1 \over 2} (B_m r_m)^2 v_{mz}~.
\end{equation}
Assuming a {\it stationary flow} with respect to the pulsar, the
synchrotron losses along the magnetotail imply
$$
{d {\cal F}_E \over dz^\prime} = -(\pi r_m^2 n_\ell) {2\over
3}r_e^2 c B_m^2 <\gamma_\ell^2>~,
$$
or
\begin{equation}
{d (\chi^2 v_{mz}) \over dz^\prime}= - {1\over z_0}\left({r_{m0}
\over r_m}\right)^2 \chi^6 ~v_{mz}~,
\end{equation}
where $z^\prime \equiv -z \geq 0$, and
$$
\chi \equiv (B_m/B_{m0})(r_m/r_{m0}) \leq 1~.
$$
Here, $z_0$ is the {\it radiation length-scale} of the
magnetotail,
$$
z_0={24 m^2 c^4 \over B_L r_L r_e^2 ~e~ (4\pi \rho_{ism} v^2)~
{\cal C_\ell}}~,
$$
\begin{equation}
\approx 15~ {P^2 \over \mu_{30}~ n_{ism}~ v_8^2 ~{\cal
C}_{\ell}}~{\rm pc}~,
\end{equation}
and ${\cal C_\ell} \equiv <\gamma_\ell^2>/<\gamma_\ell>^2$ is a
constant of order unity.
     Solutions of
equation (29) for $v_{mz}=$ const, indicate that $\chi$ decreases
very gradually with $z^\prime$.

The frequency of the peak of the synchrotron radiation from the
magnetotail is
$$
\nu_{syn} \approx {1 \over 4 \pi} { e B_m \gamma^2 \over mc}
$$
\begin{equation}
\approx 7.4\times 10^{14}~{\mu_{30}^2~ n_{ism}^{1/2}~ v_8 \over
P^4}~ \chi^5 \left({r_{m0} \over r_m}\right) \left({v_{mz} \over
10^{10}{\rm cm/s}}\right)^2~{\rm Hz}~,
\end{equation}
where we have used equation (27) for lepton Lorentz factor.

%%%%%%%%%%%%%%%%%%%%%%%%%%%%%%%%%%%%%%%%%%%%%%%%%
\begin{figure*}[t]
\epsscale{1.0} \plotone{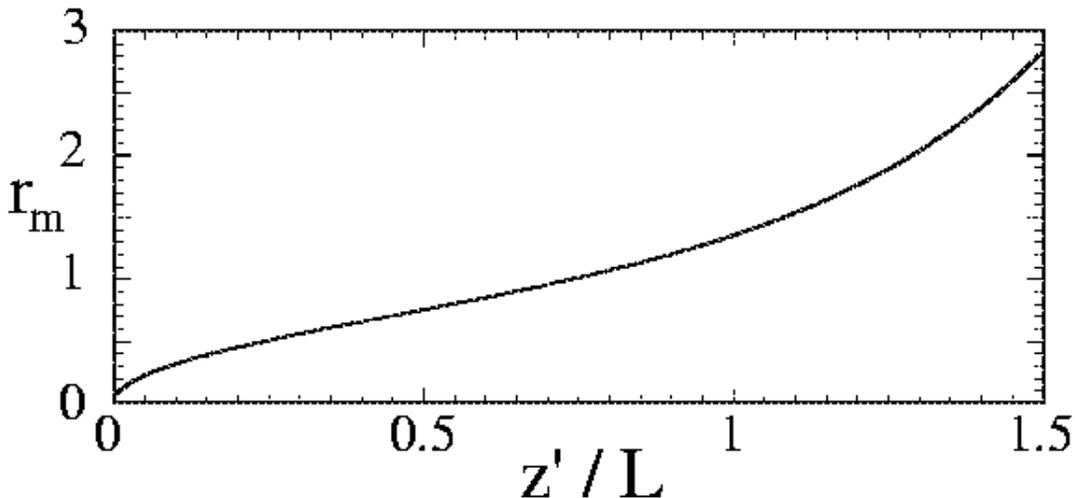} \caption{Qualitative dependence
of the width of the magnetotail $r_m(z^\prime)$ on the distance
behind the star $z^\prime$. The vertical scale is arbitrary; it
given by $(z^\prime/L)^{1/2} F(z^\prime/L)$, with $F$ given below
equation (34). }
\end{figure*}
%%%%%%%%%%%%%%%%%%%%%%%%%%%%%%%%%%%%%%%%%%%%%%%%%

     Most pulsar velocities through
the ISM are less than $10^3$ km/s, with typical velocities $\sim
100 - 500$ km/s (Arzoumanian, Chernoff, \& Cordes 2002). For these
slower pulsars, the length $z_0$ may be larger than the total
distance the pulsar has travelled. For such conditions only a
small fraction of the magnetotail energy flux ${\cal F}_E$ is lost
to synchrotron radiation. Here we discuss the long distance
$z^\prime \gg r_{m0}$ behavior of the magnetotail in this limit.
The internal pressure of the magnetotail, $p_m$, is much larger
than the pressure of the ISM at the distance $z^\prime$, $p_{ism}
=\rho_{ism} c_s^2/\Gamma \ll p_{m}$. In general the radius of the
magnetotail increases with distance from the star. Assuming a
stationary configuration in the reference frame moving with the
star we have
$$
{d r_m(z^\prime) \over d z^\prime} = {v_{mr} \over v_{mz}}~,
$$
\begin{equation}
v_{mr} =\left({p_m(z^\prime) \over \rho_{ism}}\right)^{1/2}
=\left({\dot{E}_w \over 3\pi r_m^2 v_{mz} \rho_{ism}}
\right)^{1/2}~.
\end{equation}
Thus
$$
{d r_m(z^\prime) \over d z^\prime}= \left({4c \over 3
v_{mz}}\right)^{1/2} {r_{m0} \over r_m}~ {v \over v_{mz}}~,
$$
\begin{equation}
\approx 0.02~ v_8 \left( {r_{m0} \over r_m}\right)
\left({10^{10}{\rm cm/s} \over v_{mz}}\right)^{3/2}~.
\end{equation}
Expressed in this form, $dr_m/dz^\prime$ is independent of
$\dot{E}_w$ and $\rho_{ism}$.
         Equation (32) represents a balance of the internal
pressure of the magnetotail, $p_m(z^\prime)$, against the ram
pressure  due to the tail's expansion into the ISM,
$\rho_{ism}(dr_m/dt)^2$, where $dz^\prime/dt=v_{mz}$.

For distances $z^\prime \gg r_{m0}$ but small compared with the
total distance the star has travelled, $v_{mz}$ is approximately
constant.
     Consequently,
\begin{equation}
r_m(z^\prime) \approx 0.2~ \big( r_{m0}~z^\prime\big)^{1/2}
\left({10^{10}{\rm cm/s} \over v_{mz}}\right)^{3/4}~,
\end{equation}
for $z^\prime \gg r_{m0}$.

The situation is different for distances $z^\prime$ of the order
of or somewhat larger than the total distance travelled by the
star, $L \equiv v t\approx 10 v_8 (t/10^4{\rm yr})~{\rm pc} $,
with $t$ the age of the star. The velocity of the magnetotail
plasma $v_{mz}$ must decrease strongly with $z^\prime$ of the
order of this distance owing to collisions with the plasma of the
above-mentioned bubble ${\cal R}(t)$ formed at the birth of the
pulsar. For a rough description consider
$$
v_m(z^\prime)= v_m(0)\exp\left(-{(z^\prime)^2\over L^2}\right)~.
$$
For this case equation (33) gives
\begin{equation}
r_m(z^\prime) =0.2 \big(~ r_{m0}z^\prime~\big)^{1/2}
\left({10^{10}~{\rm cm/s} \over v_{mz}(0)}\right)^{3/4}
F\left({z^\prime \over L}\right)~,
\end{equation}
where
$$
F(\xi) \equiv\left({1\over \xi} \int_0^\xi d\xi \exp\left({3\xi^2/
2} \right)\right)^{1/2}~.
$$
Figure 4 shows the qualitative dependence of $r_m(z^\prime)$.
Notice that the curvature of the magnetotail radius $d^2
r_m/dz^{\prime~\!2}$ is initially negative but at large $z^\prime
\gtrsim 0.45 L$ it becomes positive. Further note that
$v_{mz}(z^\prime) [r_m(z^\prime)]^2$ is an increasing function so
that $B_m(z^\prime)$ decreases with distance from the star.

The magnetotails may be observable as low surface brightness
regions of non-thermal, polarized (synchrotron) emission.

\section{Examples of Magnetotails: Guitar and Mouse Nebulae}

     The Guitar Nebula is  created by the high velocity
pulsar B2224+65. Chatterjee and Cordes (2002, 2004) estimate the
velocity of the pulsar as $v_8 \approx 1.7$ or $1700$ km/s based
on a distance to the source of $D=1.9$ kpc.
      The pulsar period $P=0.68$ s
and spin-down rate $\dot{P}=9.7\times 10^{-15}$s/s imply a power
output $\dot{E} \approx 1.2\times 10^{33}$ erg/s assuming a moment
of inertia $I=10^{45}$ g cm$^2$.
     The angular length of the
tail is $\delta \theta \approx 15^{''}$ (Chatterjee \& Cordes
2004) which corresponds to a tail length $z^\prime \approx
0.15~{\rm pc}/\sin\iota$, where $\iota$ is the angle between the
jet axis and the line of sight.
      With $\dot{E}=B_L^2 r_L^2 c$,
this power implies $\mu_{30} \approx 2.1$.
      Consequently, equation (30) gives the length-scale
of the magnetotail $z_0 \approx 1~{\rm pc}/(n_{ism}{\cal
C}_\ell)$.
%%%%%%%%%%%%%%%%%%%%%%%%%%RRR%%%%%%%%%%%%%%%%%%%%%%%%%%%%%%%%
This corresponds approximately  to the whole length of the tail in
the Guitar nebula observed in the $H_\alpha$ line.
%%%%%%%%%%%%%%%%%%%%%%%%%%%%%%%%%%%%%%%%%%%%%%%%%%%%%%%%%%%%%%%%
     Equation (31) gives the estimate
$\nu_{syn} \approx 2.6\times 10^{16}~{\rm Hz}~ n_{ism}^{1/2}~
\chi^5(r_{m0}/r_m)(v_{mz}/10^{10} {\rm cm/s})^2$.
%%%%%%%%%%%%%%%%%%RRR%%%%%%%%%%%%%%%%%%%%%%%
The Guitar nebula was observed as a source of very weak radiation
in soft X-rays which is a sign of the presence of highly
relativistic leptons, $\gamma \sim 10^7$. (Romani, Cordes, \&
Yadigaroglu 1997).
    Equation (13) also gives similar $\gamma$
for the Guitar nebula.
     Both, the magnetic field and the particles interact
with the interstellar medium and are deflected to the tail.
   The radio images of the Guitar nebula show clear limb-
brightening (e.g., Chatterjee \& Cordes 2004).
    This may be due to the
compression of the magnetic field at the shock wave which  leads
to higher intensity of radiation (e.g., Romani et al. 1997).
   From other side,  enhanced reconnection is expected
near the shock wave due to compression
the magnetic field components of
opposite polarity.
    This will also lead
to  limb-brightening.
    The inner (smaller scale) structure
of the Guitar nebula is closed in the back and may represent the
inner shock wave, where  accelerated particles were stopped by the
interstellar medium. If this is the case, then the flow has
$\sigma < 1$ before the shock wave.

Another example is a Mouse Nebula which is created by much more
powerful pulsar PSR J1747-2958 which propagates with high velocity
$v\approx 600~{\rm km/s}$ (Camilo et al. 2002). It has a period
$P=0.098$ s  and a spin-down luminosity $\dot{E}\approx 2.5\times
10^{36}$ erg/s which corresponds to $\mu_{30}\approx 2.5$.
    The extended radio nebulae was
discovered around this pulsar (Yusef-Zadeh \& Bally 1987).
    An interesting feature is the tail which propagates to very far
distances. Recently,  an X-ray nebula was found in observations
with Chandra (Gaensler et al. 2004).
   Gaensler et al. (2004) estimate the X-ray
luminosity of the nebula as $L_X(0.5-8~{\rm keV})\approx 5\times
10^{34}$ erg/s based on a distance to the source of $D=5$ kpc, the
angular length of the Mouse tail as $\delta \theta \approx
45^{''}$ which corresponds to a tail length
   $z^\prime \approx 1~{\rm pc}/\sin\iota$.
      For these parameters and assuming $v_8=0.6$
and $n_{ism}=0.3$/cm$^3$ (Gaensler et al. 2004), equation (30)
gives radiation length-scale of the magnetotail, $z_0 \approx
0.5{~\rm pc}/{\rm C}_\ell$.
        For this object equation (31) gives the estimate
$\nu_{syn} \approx 2.7\times 10^{19}~{\rm Hz}~ n_{ism}^{1/2}~
\chi^5(r_{m0}/r_m)(v_{mz}/10^{10} {\rm cm/s})^2$.
    The predicted
frequency and overall luminosity is much higher than those for
Guitar nebula, which correspond to observations of much brighter
X-ray source.

The shape of the Mouse nebula was modelled by hydrodynamic
simulations (Gaensler et al. 2004; van der Swaluw et al. 2003; and
Bucciantini 2002).
    It was suggested
that the inner brightest region of the nebula may be due to
the termination shock of the pulsar wind, while the long tail may
represent the post-shock flow.
%%%%%%%%%%%%%%%RRR%%%%%%%%%%%%%%%%%%%%
We note that the  Chandra observations do now show evidence of
the shock wave in the X-ray picture of this nebula.
   Instead, the
X-ray luminosity in the brightest region (the head of the Mouse)
varies gradually (Gaensler et al. 2004).  This may be a sign that
in the tail the $\sigma$ parameters is not small and the back flow
is not super-Alfv\'enic.
    If $\sigma  \sim 1$ in the tail, then the
magnetotail may propagate to very far distances (see, e.g.,
simulations by Romanova et al. (2001)
and Toropina et al. (2001).
  Radio radiation is then connected with particles which propagate from
the pulsar along the tail to far distances and lose their energy
there, or, particles may be accelerated in situ during the
reconnection processes in the magnetotail (Toropina et al. 2001).

The direction of jets observed in a few cases is almost aligned
with the direction of propagation of the pulsar and seems to be
perpendicular to the PWN disk (e.g., Ng \& Romani 1987).
   If the direction of the pulsar's
motion is determined by a magnetic
kick during its formation (Ardeljan, Bisnovatyi-Kogan
\& Moiseenko 2001; Lai, Chernoff \&
Cordes 2000) then both, $\rvecomega_*$
and $\rvecmu$ will be in the same
direction.
    If the
magnetic axis is misaligned
relative to the rotational axis, but
$\rvecomega_*$ is aligned with
the pulsar's velocity  ${\bf
v}$, then the situation is
similar to the one described above, because the
jet forms in the direction of
$\rvecomega_*$ and the disk magnetic
structure is perpendicular to $\omega_*$.
    However, if $\rvecomega_*$
does not coincide with ${\bf v}$,
then both the jet and the PWN
disk may give non-axisymmetric features
of the pulsar wind nebula.
   This later situation has not been
investigated.
   A related situation appears in the recently
discovered double pulsar
binary PSR J0737-3039 where the direction of
the flow from the wind-generating
pulsar changes in time because
of the motion of stars in the
binary system (Kaspi et al. 2004; Demorest
et al. 2004).
    Three dimensional simulations
of such a flow have shown that for
large inclination angles between magnetic and rotational axes of
pulsar, the shock wave may be non-axisymmetric (Spitkovsky \&
Arons 2004).

\section{Observability of Wind and Magnetotail
Emitting Stars in the Solar Neighborhood}

Although neutron stars which spin-down
to periods $P \gtrsim 1-3~{\rm
s}$ shut off as pulsars, they may continue to emit a wind and
interact with the ISM. The number of such objects is larger than
the number of short period pulsars. It is possible that such
objects are detectable at distances $\lesssim 10^2$ pc.
Interesting nearby isolated neutron star candidate RX J1856.5-3754
has a prominent shock wave, and we discuss this object below.

Different spin-down powers $\dot{E}_w$ are expected depending on
the star's period of rotation and magnetic field. We consider two
cases, one with a relatively high spin-down power $\dot{E}_w\sim
10^{31} - 10^{32}~{\rm ergs/s}$ which is pertinent to old pulsars,
or relatively powerful shutoff pulsars, and a low spin-down power,
$\dot{E}_w =10^{28} {\rm ergs/s}$ corresponding to very weak
shutoff pulsars.

\subsection{Relatively High Spin-Down Power Objects}

Shutoff or misdirected pulsars
with periods $P \sim 1~{\rm s}$ and
a surface magnetic field $B \sim 10^{12}~{\rm G}$ have a
relatively high spin-down power, $\dot{E}_w \sim 6\times 10^{31}$
erg/s (i.e., equation 2).
         The synchrotron emission
from the pulsar wind for distances within the bow shock wave is
given by equation (20) and the photon energies are given by
equation (22) and are in the high X-ray range.
        Additionally, there will
be a small power in thermal radiation from the shocked ISM
according to equation (5).
       For high velocity pulsars,
say $v > 500$ km/s, a significant fraction of the spin-down power
is emitted in the magnetotail.
       The photon energies
are then much lower, in the UV band
according to equation (31).
    Such shut-off or misdirected pulsars
may be observed owing to the radiation of their tails.

\subsection{Nature of RX J1856.5-3754}

The source RX J1856.5-3754 is
an interesting isolated neutron star
candidate located at a
distance $d\approx60~{\rm pc}$ (Walter
2001). It has a bow shock wave observed by van Kerkwijk and
Kulkarni (2001a,b) in the ${\rm H}_\alpha$ line.
The observed
standoff distance between the star and the bow shock wave is
$\approx 1''$ which corresponds to $ z_{sh} \approx 0.9
\times10^{15}~{\rm cm}$. The bow shock is probably due to the
interaction of star's wind with the ISM (see, e.g., van Kerkwijk
and Kulkarni 2001b). In this case we obtain from equation (3) the
relation $B_{12}/(n_{ism}^{1/2} v_8 P^2) \approx 9$, where
$B_{12}$ may be replaced by $\mu_{30}$. This relation implies that
the spin-down power of the star is $\dot{E}_w \approx 5.2 \times
10^{32} n_{ism} v_8^2$ erg/s $\approx 5.2\times 10^{30}$ erg/s,
where we took $v_8=0.1$ and $n_{ism}=1/{\rm cm^3}$.

Clearly, a range of values of $B_{12}$ (or $\mu_{30}$), $P$, and
$v_8$ can give $\mu_{30}/(v_8 P^2) \approx 9$, where we have taken
for simplicity $n_{ism} = 1$ ($1/{\rm cm}^3$). To consider the
likely values, note that the spin-down time of the star is $t_{sd}
\equiv P/\dot{P} = I P^2 c^3/(4\pi^2 \mu^2) \approx 2.2 \times
10^7{\rm yr} ~\!P^2/\mu_{30}^2$, assuming $I=10^{45}$ g cm$^2$.
The total number of neutron stars likely to be within a distance
of $60$ pc of the Sun is $N \approx 10^9(60{\rm pc}/10^4{\rm
pc})^3 \approx 220$ (e.g., Schvartsman 1978; Treves \& Colpi
1991). Then the probable number of objects with spin-down time
$t_{sd}$ within $60$ pc is $N^\prime \approx 220(t_{sd}/t_{H})$,
where $t_{H}\approx1.5\times10^{10}~{\rm years}$ is the Hubble
time. Thus, for the object RX J1856.5-3754, $N^\prime \approx
0.036/(\mu_{30} v_8)$. The corresponding rotation period of the
star is $P\approx 0.33(\mu_{30}/v_8)^{1/2}$ s. Clearly, larger
values of $N^\prime$ occur for a smaller values of the product
$\mu_{30} v_8$. For example, for $\mu_{30}=0.1$ and $v_8=0.1$,
$N^\prime \approx 3.6$. For these values $\dot{E}_w \approx
5.2\times 10^{30}$ erg/s. So, this object may be a misdirected
pulsar if magnetic field is large, or, shutoff pulsar if the field
is small.

\subsection{Low Spin-Down Power
Objects}

There should be a much larger number of shutoff pulsars with the
lower spin-down power.
   As an example, we take objects with the
spin-down power $\dot{E}_w=10^{28}$ erg/s.
   For these objects,
$\mu_{30}^2/P^{4}\approx 1.7\times 10^{-4}$, so that $P \approx
8.7 \mu_{30}^{1/2}~{\rm s}$.
   The number of such objects within a
distance of $60$ pc of the Sun is $N^\prime \approx
220(t_{sd}/t_H) \approx 24/\mu_{30}$.
    The standoff distance of the
bow shock wave is however small, $z_{sh} \approx 1.3 \times
10^{12}~ v_8^{-1}$ cm, but much larger than the light cylinder
radius.

\section{Discussions and Conclusions}

This work considers rotating
magnetized neutron stars emitting
  power $\dot{E}_w$ in a wind of relativistic particles
and electromagnetic field.
We argue that a non-negligible fraction
of the wind's power may be converted to relativistic particles due
to annihilation or reconnection of the magnetic field. Outside of
the light cylinder, the star's rotation acts to wind up the
magnetic field to form a predominantly azimuthal, slowly
decreasing with distance, magnetic field of opposite polarity on
either side of the equatorial plane normal to the star's rotation
axis. An analogous situation exists in the Solar Wind (Bertin \&
Coppi 1985). The magnetic field annihilates across the equatorial
plane with the magnetic energy going to accelerate the charged
particles to highly relativistic energies. The accelerated leptons
emit synchrotron radiation in a broad range from the UV to gamma
ray energies. Additionally, there may be oppositely directed
Poynting outflows along the star's rotation
axis. These outflows are analogous
to the Poynting jets discussed by
Lovelace and Romanova (2003).
        The model is qualitative in  the respect
that we do not have global self-consistent calculations of the
electromagnetic fields and sources.

For a typical, supersonically moving star, the star's relativistic
wind forms a bow shock wave with the interstellar medium
(Schvartsman 1970). We argue that an appreciable fraction of the
star's wind power $\dot{E}_w$ is deflected by the bow shock wave
and collimated into the star's magnetotail.
    Plasma moves down the
magnetotail with a relativistic velocity.
   Further, we argue that
equipartition is reached in the magnetotail between the magnetic
energy and the relativistic particle energy. For the case where
the charged particles are leptons, the synchrotron radiation
length-scale $z_0$ of the magnetotail is calculated. Over this
distance the energy flux in the magnetotail decreases by a factor
of order $2$. For highly supersonic pulsars, $z_0$ may be less
than the total distance the star has travelled.
     The synchrotron
radiation spectrum is expected to be nonthermal with typical
photon energies in the UV range for
the highly supersonic pulsar B2224+65 which generates a bow shock
wave and the Guitar Nebula (Chatterjee \& Cordes 2002).

The ratio of the power due to the shock heating of the ISM, which
gives thermal emission, to the wind power $\dot{E}_w$, which gives
nonthermal synchrotron emission, is shown to be $(3/64)(v/c) \ll
1$, where $v$ is the velocity of the star and $c$ the speed of
light.

An equation is derived for
the radius of the magnetotail
$r_m(z^\prime)$ as a function
of distance $z^\prime$ from the
star. For large distances $z^\prime$,
of the order of the distance
travelled by the star,
we argue that the magnetotail has a
`trumpet' shape owing to the
slowing down of the magnetotail flow.

We argue that the isolated neutron
star candidate RX J1956.5-3754
may be a misdirected or shutoff pulsar.

We estimate the number of shutoff pulsars which may be observable
in the vicinity of the Sun for cases of
relatively strong ($10^{32}$ erg/s)
and weak ($10^{28}$ erg/s) power.
   A much larger number of weak
shutoff pulsars is predicted.

{We thank Drs. Kaya Mori,  David Chernoff, and Jim Cordes for
stimulating discussions.
   We thank an anonymous referee for
thoughtful comments.
This work was supported in part by NASA
grants NAG 5-13060, NAG 5-13220, and NSF grant AST-0307817.}

\end{document}